\documentclass[12pt]{article}
 \usepackage[bookmarks=false]{hyperref}
\usepackage{times}
\usepackage{amsmath}
\usepackage{amsfonts}
\usepackage{amssymb}
\usepackage{graphicx}
\usepackage[natbibapa]{apacite}  
\usepackage{tikz}
\usetikzlibrary{arrows.meta,positioning}
\usepackage{pgfplots}
\pgfplotsset{compat=1.18}

\advance\topmargin by -0.6truein
\newcommand{\be}{\begin{eqnarray}}
\newcommand{\ee}{\end{eqnarray}}
\newcommand{\ba}{\begin{eqnarray*}}
\newcommand{\ea}{\end{eqnarray*}}

\newcommand{\dom}{\preceq}
\newcommand{\rdom}{\succeq}
\newcommand{\asym}{\asymp}

\begin{document}
\setcounter{figure}{0}

\begin{center}
{\Large \bf Adjusted Kolmogorov Complexity of Binary Words with Empirical Entropy Normalization}
\end{center}

\begin{center}
    {\large Brani Vidakovic}\\
    Department of Statistics, Texas A\&M University
\end{center}

\vspace{0.5in}

\begin{quote}
{\footnotesize
Kolmogorov complexity of a finite binary word reflects both algorithmic structure
and the empirical distribution of symbols appearing in the word. Words with
symbol frequencies far from one half have smaller combinatorial richness and
therefore appear less complex under the standard definition. In this paper an
entropy-normalized complexity measure is introduced that divides the Kolmogorov
complexity of a word by the empirical entropy of its observed distribution of
zeros and ones. This adjustment isolates intrinsic descriptive complexity from
the purely combinatorial effect of symbol imbalance. For Martin L\"of random
sequences under constructive exchangeable measures, the adjusted complexity
grows linearly and converges to one. A pathological construction shows that
regularity of the underlying measure is essential. The proposed framework
connects Kolmogorov complexity, empirical entropy, and randomness in a natural
manner and suggests applications in randomness testing and in the analysis of
structured binary data.


}
\end{quote}

\section{Introduction}
\label{sec:intro}

Kolmogorov complexity, introduced in the seminal works of
\citet{Kolmogorov1965}, \citet{ZvonkinLevin1970}, and \citet{Levin1974}, provides
an algorithmic notion of complexity for finite objects and infinite sequences.
A binary word is regarded as complex if the length of the shortest program that
outputs it on a universal prefix machine is essentially equal to the word
length itself, whereas highly structured words admit substantially shorter
descriptions.

A classical and well-known observation is that Kolmogorov complexity is
strongly influenced by the empirical distribution of symbols. If a word of
length $n$ contains $n_1$ ones, then the logarithmic size of its combinatorial
class is approximately $nH(p)$, where $p=n_1/n$ and $H$ denotes binary entropy.
As $H(p)$ varies with symbol frequencies, raw Kolmogorov complexity does not
fully separate intrinsic descriptive structure from the combinatorial effect of
symbol imbalance.

To address this issue, we analyze an entropy-normalized complexity scale
\be
KA(x) = \frac{K(x)}{H(\nu_{\ell(x)})},
\label{eq:KA_def_intro}
\ee
where $K(x)$ denotes the Kolmogorov complexity of a finite binary word $x$ and
$H(\nu_{\ell(x)})$ is the empirical entropy of its observed distribution of zeros
and ones. The normalization is undefined for constant words (zero empirical
entropy) but is meaningful for all other finite words. 
This normalization was first indicated in \citet{ZvonkinLevin1970}, but without deeper analysis.

The main theorem of the
paper shows that for Martin L\"of random sequences under constructive
exchangeable measures, the ratio $KA(\omega^m)/m$ converges to one. In this
sense, empirical entropy provides the correct normalization scale for the
descriptive complexity of random sequences.

This entropy-normalized formulation builds on earlier work in which algorithmic
complexity, Martin L\"of tests, universal priors, and Ockham-type principles
were developed within a unified framework, with particular emphasis on
measures on $\Omega$ and on Schnorr-style notions of complexity
\citep{ZvonkinLevin1970,Schnorr1971,Vidakovic1988}. The conceptual distinction
between combinatorial entropy and descriptive complexity has been emphasized
repeatedly in the algorithmic information literature; see, for example,
\citet{Kramosil1982} and \citet{Calude2002}, where information content is treated
as an intrinsic structural property not reducible to symbol frequencies alone.

The present normalization isolates the role of empirical entropy at the level of
\emph{finite} words, where the combinatorial effect of imbalance is explicit and
quantifiable before any asymptotic limit is taken. This finite-sample viewpoint
becomes particularly important later in the paper, when we develop effective
surrogates and implementable randomness tests in
Section~\ref{sec:effective_testing}.

The work is also closely related to earlier attempts to disentangle algorithmic
structure from combinatorial effects in binary sequences. In
\citet{Vidakovic1983Process}, a process-based conditional complexity (KP
complexity) was introduced together with an information measure defined through
differences of conditional complexities. That framework emphasized the role of
conditioning objects and generating processes in quantifying information and
randomness, and it established asymptotic links between complexity growth rates
and Schnorr- and Martin L\"of-type randomness.

From a complementary perspective, an effective combinational measure of
complexity for binary words was studied in
\citet{StojanovicVidakovic1987}. There it was shown that complexity is strongly
influenced by word weight and that substantial deviations from balance
necessarily reduce achievable complexity. That work also established a
Shannon-type concentration phenomenon: for a fixed length, most words have
complexity close to the maximum permitted by their combinatorial class.

The entropy-normalized Kolmogorov complexity introduced here may be viewed as an
algorithmic counterpart of these ideas. By explicitly factoring out the
combinatorial contribution of symbol imbalance, the normalization isolates
intrinsic descriptive complexity and operates directly within the framework of
Kolmogorov complexity and algorithmic randomness. Moreover, the approach extends
naturally to conditional and mutual settings (see
Section~\ref{sec:conditional_mutual}), thereby providing a unifying bridge
between effective combinational measures, process-based conditional
complexities, and entropy rates arising in information theory.

\section{Roadmap and Contributions}
\label{sec:roadmap}

This paper develops an empirical-entropy normalization of Kolmogorov complexity
for finite binary words and relates the resulting normalized quantities to
Martin--L\"of randomness under regular (constructive) probability measures on
$\Omega=\{0,1\}^{\mathbb{N}}$.

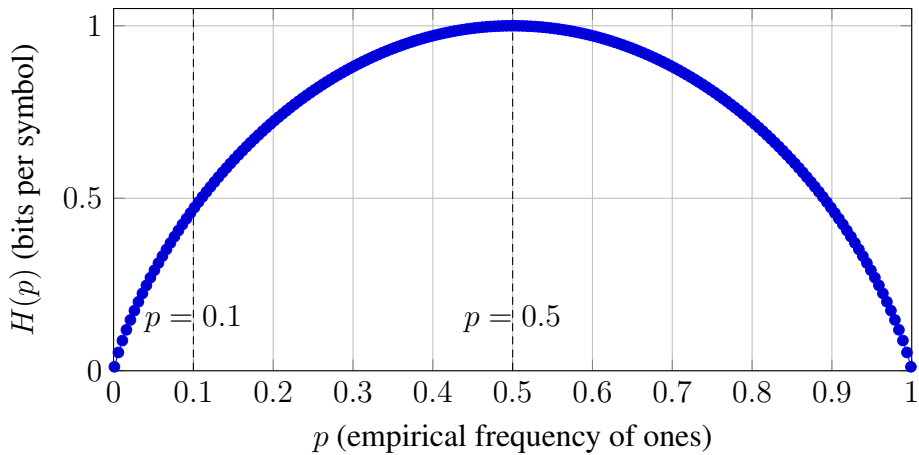
\begin{figure}[htb]
\centering
\begin{tikzpicture}
\begin{axis}[
  width=0.80\textwidth,
  height=0.42\textwidth,
  xlabel={$p$ (empirical frequency of ones)},
  ylabel={$H(p)$ (bits per symbol)},
  xmin=0, xmax=1,
  ymin=0, ymax=1.05,
  samples=200,
  domain=0.001:0.999,
  grid=both,
]
\addplot {-(x*log2(x) + (1-x)*log2(1-x))};
\addplot[densely dashed] coordinates {(0.1,0) (0.1,1.05)};
\addplot[densely dashed] coordinates {(0.5,0) (0.5,1.05)};
\node at (axis cs:0.10,0.15) {$p=0.1$};
\node at (axis cs:0.50,0.15) {$p=0.5$};
\end{axis}
\end{tikzpicture}
\caption{\small Binary empirical entropy $H(p)$ as a function of symbol imbalance. The
combinatorial baseline for words with empirical frequency $p$ is $mH(p)$ rather
than $m$.}
\label{fig:entropy_curve1}
\end{figure}

The exposition is organized around a single central ratio,
\be
R(x)=\frac{K(x)}{\ell(x)\,H(\nu_{\ell(x)})},
\label{eq:R_def_roadmap}
\ee
which compares the Kolmogorov complexity of a word to the empirical
combinatorial baseline determined by its symbol frequencies. The quantity
$KA(x)$ in \eqref{eq:KA_def_intro} is simply the corresponding scale,
$KA(x)=\ell(x)R(x)$. Conceptually, $R(x)$ is the primary statistic: both the
numerator and denominator are measured in bits, and the ratio directly
quantifies deviation from shell-typical behavior. The empirical entropy
baseline is introduced formally in
Section~\ref{sec:KA} and visualized in
Figure~\ref{fig:entropy_curve1}.

\subsection*{Notation convention used throughout}
\label{sec:notation}

We maintain a strict separation between notation for finite words and for
prefixes of infinite sequences, and we adopt a consistent notation for empirical
frequencies.

If $x$ denotes a finite binary word, then $\ell(x)$ denotes its length and
$w(x)$ its weight, that is, the number of ones appearing in $x$. The empirical
proportion of ones is
$$
p(x)=\frac{w(x)}{\ell(x)},
$$
and the associated empirical distribution is
$$
\nu_{\ell(x)}=(p(x),1-p(x)).
$$

If $\omega\in\Omega$ denotes an infinite binary sequence, then $\omega^m$
denotes its prefix of length $m$. The weight of the prefix is $w(\omega^m)$,
the empirical proportion of ones is
$$
p_m(\omega)=\frac{w(\omega^m)}{m},
$$
and the corresponding empirical distribution is denoted by $\nu_m(\omega)$.
Throughout, $m$ refers exclusively to prefix length, while $\ell(x)$ denotes
the length of a finite word.

\subsection*{Adjusted complexity and the normalization principle}

A classical counting argument shows that for fixed integers $(m,k)$ the shell
\[
S(m,k)=\{x\in\{0,1\}^m:w(x)=k\}
\]
has cardinality $\binom{m}{k}$. Consequently, the natural combinatorial baseline
for describing a shell-typical word is
\begin{eqnarray}
\log_2|S(m,k)|
&=&
\log_2\binom{m}{k}
\nonumber\\
&=&
mH(k/m)+O(\log m).
\label{eq:shell_size_roadmap}
\end{eqnarray}
This observation motivates the entropy-based normalization that factors out
symbol imbalance from intrinsic algorithmic structure.

For a finite word $x$ with $H(\nu_{\ell(x)})>0$, we define the normalized
complexity ratio by \eqref{eq:R_def_roadmap} and the corresponding adjusted
complexity scale by
\be
KA(x)=\ell(x)\,R(x)=\frac{K(x)}{H(\nu_{\ell(x)})}.
\label{eq:KA_def_roadmap}
\ee
Values of $R(x)$ near one indicate that the descriptive cost of $x$ matches the
shell baseline implied by its empirical distribution, whereas values
substantially below one indicate algorithmic regularity beyond what symbol
frequencies alone can explain. This intuition is sharpened into a precise
counting statement in Section~\ref{sec:KA} and becomes the key step in the
Martin--L\"of test construction in
Section~\ref{sec:KA_theorem_proof}.

\subsection*{Contributions}

The paper makes four main contributions.

\noindent
{\bf (i) Empirical-entropy normalization for finite words.}
We formalize the empirical entropy baseline for binary words and derive sharp
shell bounds showing that, for a fixed empirical distribution, only an
exponentially small fraction of words can exhibit a given level of complexity
deficiency. 

\noindent
{\bf (ii) Limit theorem for Martin L\"of random sequences.}
For an infinite sequence $\omega$ that is Martin L\"of random with respect to a
regular constructive measure $\mu$, we establish the asymptotic identity
\begin{eqnarray}
\frac{K(\omega^m)}{m}
&\longrightarrow&
h(\mu),
\nonumber\\
R(\omega^m)
&=&
\frac{K(\omega^m)}{m\,H(\nu_m(\omega))}
\longrightarrow
1,
\label{eq:main_limit_form_roadmap}
\end{eqnarray}
where $h(\mu)$ is the entropy rate of the measure. The proof appears in
Section~\ref{sec:KA_theorem_proof}, followed immediately by a discussion of
regularity in Section~\ref{sec:pathological}.

\noindent
{\bf (iii) Necessity of regularity assumptions.}
We construct a pathological counterexample showing that without appropriate
regularity conditions on $\mu$, empirical entropy may fail to provide the
correct normalization scale, motivating higher-order empirical summaries.

\begin{figure}[t]
\centering
\begin{tikzpicture}
\begin{axis}[
  width=0.78\textwidth,
  height=0.28\textwidth,
  xlabel={$R_{\mathrm{eff}}(x) = K_{\mathrm{eff}}(x)/(nH(\nu_n))$},
  xmin=0, xmax=1.4,
  ymin=0, ymax=1,
  axis y line=none,
  ytick=\empty,
  grid=both,
  clip=false
]
\addplot[densely dashed] coordinates {(0.83,0) (0.83,1)};
\node[anchor=south] at (axis cs:0.83,0.95)
  {$c(m)=1-\dfrac{m}{nH(\nu_n)}$};
\node[anchor=west] at (axis cs:0.15,0.55)
  {Reject (structure)};
\node[anchor=west] at (axis cs:0.85,0.55)
  {Accept (shell-typical)};
\end{axis}
\end{tikzpicture}
\caption{\small Toy effective randomness test expressed in the normalized ratio
$R_{\mathrm{eff}}(x)$. For a fixed deficiency threshold $m$, randomness is
rejected whenever $R_{\mathrm{eff}}(x)\le c(m)=1-m/(nH(\nu_n))$.}
\label{fig:toy_test}
\end{figure}
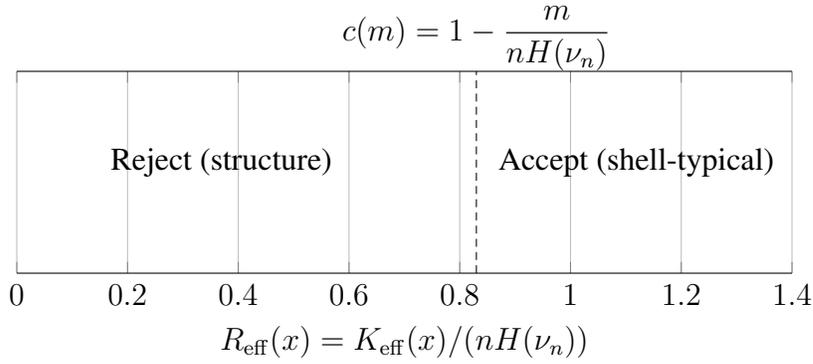

\noindent
{\bf (iv) Effective surrogates and implementable tests.}
By replacing Kolmogorov complexity with computable upper bounds, we obtain
effective statistics
\[
R_{\mathrm{eff}}(x)=
\frac{K_{\mathrm{eff}}(x)}{\ell(x)\,H(\nu_{\ell(x)})},
\]
leading to operational randomness tests developed in
Section~\ref{sec:effective_testing} and illustrated in
Figure~\ref{fig:toy_test}.

\medskip

The paper is organized as follows.
Section~\ref{sec:preliminaries} introduces notation and preliminaries.
Section~\ref{sec:KA} develops the basic properties of $R(x)$ and $KA(x)$.
Section~\ref{sec:KA_theorem_proof} proves the main limit theorem, and
Section~\ref{sec:pathological} explains why regularity is essential.
Sections~\ref{sec:algo_interp} and \ref{sec:stat_effective_uses} develop
algorithmic and applied interpretations.
Section~\ref{sec:conditional_mutual} treats conditional and mutual extensions,
and Section~\ref{sec:effective_testing} consolidates effective testing logic.
Section~\ref{sec:discussion} concludes.

\section{Preliminaries}
\label{sec:preliminaries}


Let $\Omega=\{0,1\}^{\mathbb{N}}$ denote the space of all infinite binary
sequences. For $\omega\in\Omega$, its prefix of length $m$ is written
$\omega^{m}=\omega_1\omega_2\cdots\omega_m$. For a finite binary word
$x=x_1x_2\cdots x_{\ell(x)}$, the associated cylinder set is
\be
\Gamma_x=\{\omega\in\Omega:\omega^{\ell(x)}=x\}.
\label{eq:cylinder_def}
\ee
Cylinder sets generate the Borel $\sigma$--algebra on $\Omega$.

We identify finite words with nonnegative integers via the standard bijection
\be
x \;\mapsto\; 2^{\ell(x)}-1+\sum_{i=1}^{\ell(x)} x_i\,2^{\ell(x)-i}.
\label{eq:word_integer_bijection}
\ee
When no confusion arises, the same symbol denotes a word and its corresponding
integer. This convention is consistent with earlier usage in
\citep{Vidakovic1983Process}.

Throughout, we use the standard domination relations up to additive constants.
For functions $a(\cdot)$ and $b(\cdot)$ on a common domain, we write
\be
a \dom b
\ee
if there exists a constant $C$ such that $a(u)\le b(u)+C$ for all relevant $u$.
The reverse relation is denoted by $a \rdom b$, and $a \asym b$ denotes mutual
domination. In the sequel we state comparisons almost exclusively in the form
$\dom,\rdom,\asym$, and we avoid rewriting them as explicit inequalities.
These relations appear frequently in algorithmic information theory and are used
explicitly in \citep{Vidakovic1983Process, LiVitanyi2008, DowneyHirschfeldt2010, Nies2009}.

\medskip


Let $\mathcal{P}$ denote the class of all partial recursive functions acting on
finite binary strings. For $F\in\mathcal{P}$, the Kolmogorov complexity of a
finite word $x$ relative to $F$ is
\be
K_F(x)=
\left\{
\begin{array}{ll}
\min\{\ell(p):F(p)=x\}, & \text{if such } p \text{ exists},\\
\infty, & \text{otherwise}.
\end{array}
\right.
\label{eq:KF_def}
\ee
There exists a universal $F_0\in\mathcal{P}$ such that for all $G\in\mathcal{P}$,
\be
K(x)\equiv K_{F_0}(x)\dom K_G(x).
\label{eq:universal_K}
\ee
We interpret $K(x)$ as a description length in bits.

Monographs by \citep{LiVitanyi2008, DowneyHirschfeldt2010, Nies2009} provide
comprehensive accounts of Kolmogorov complexity and its core properties.

\medskip


Let $\mu$ be a probability measure on $\Omega$. A function $V$ defined on finite
words is called a Martin L\"of test with respect to a constructive measure $\mu$
if there exists a recursive function $g(m)\downarrow 0$ such that, for every
integer $m\ge 1$,
\be
\mu\Big(\big\{\omega\in\Omega:\sup_{r\ge 1} V(\omega^{r})\ge m\big\}\Big)\le g(m).
\label{eq:ML_test_def}
\ee
There exists a universal Martin L\"of test $U$ satisfying $U(x)\dom V(x)$ for all
tests $V$ \citep{MartinLof1966}.

A sequence $\omega\in\Omega$ is \emph{Martin L\"of random with respect to $\mu$} if
\be
\forall \text{ Martin L\"of tests } V,\qquad
\sup_{r\ge 1} V(\omega^{r})<\infty.
\label{eq:ML_random_def}
\ee
Equivalently, if $U$ is universal, then
\be
\omega \text{ is Martin L\"of random w.r.t. } \mu
\quad\Longleftrightarrow\quad
\sup_{r\ge 1} U(\omega^{r})<\infty.
\label{eq:ML_random_universal}
\ee

\medskip


If $x$ is a finite binary word, we write $w(x)$ for its weight, that is, the
number of ones occurring in $x$. Define
\be
p(x)=\frac{w(x)}{\ell(x)}, \qquad q(x)=1-p(x), \qquad \nu_{\ell(x)}=(p(x),q(x)).
\label{eq:empirical_dist_finite}
\ee
If $\omega\in\Omega$, then $w(\omega^{m})$ denotes the number of ones in the
prefix $\omega^{m}$ and
\be
p_m(\omega)=\frac{w(\omega^{m})}{m}, \qquad \nu_m(\omega)=(p_m(\omega),1-p_m(\omega)).
\label{eq:empirical_dist_prefix}
\ee

We use the binary entropy, measured in bits per symbol,
\be
H(\nu_{\ell(x)})=-p(x)\log_2 p(x)-q(x)\log_2 q(x),
\label{eq:binary_entropy_def}
\ee
and similarly $H(\nu_m(\omega))$ for prefixes. The quantity
$\ell(x)H(\nu_{\ell(x)})$ represents the logarithmic size of the type class
(shell) of words having the same empirical distribution as $x$, up to
lower-order terms. This is the quantitative reason why $\ell(x)H(\nu_{\ell(x)})$
is the correct scale to compare against $K(x)$, and it is exactly the scale used
in the ratio $R(x)$ in \eqref{eq:R_def_roadmap}.

Since both $K(x)$ and $\ell(x)H(\nu_{\ell(x)})$ are measured in bits, ratios such as
$K(x)/(\ell(x)H(\nu_{\ell(x)}))$ are dimensionally consistent. Changing the
logarithm base rescales both quantities by the same constant and does not affect
comparisons stated with $\dom,\rdom,\asym$.

\subsection{Calculable measures, semimeasures, and universal priors}
\label{sec:prelim_measures}

The main theorem requires mild regularity of the underlying measure on $\Omega$.
We adopt the framework of calculable and semi calculable measures developed in
\citet{Vidakovic1981Thesis, Vidakovic1984MartinLof, Vidakovic1988}.

A probability semimeasure $\mu$ on the algebra generated by cylinders is called
\emph{calculable} if there exists a recursive procedure that, given a finite word
$x$, approximates $\mu(\Gamma_x)$ to arbitrary precision. Semi calculable measures
admit effective upper and lower rational bounds with computable convergence rates.
Bernoulli and general i.i.d.\ measures with computable parameters belong to this
class.

Transformations acting on $\Omega$ that map calculable measures to calculable
measures induce a natural partial order on semi calculable measures and lead to
the notion of a universal prior. Informally, among all semi calculable measures
on $\Omega$, a universal prior dominates all others up to a multiplicative
constant and is therefore the least informative choice consistent with effective
approximability. In the present paper, empirical-entropy normalization plays a
parallel role at the level of finite words: it factors out the combinatorial
contribution of symbol imbalance and leaves a residual descriptive component
captured by Kolmogorov complexity. This perspective will be used again in
Section~\ref{sec:algo_interp} when we interpret $R(x)$ in the language of
universal coding and MDL.

Finally, we note that process-based versions of complexity and information
provide an additional viewpoint on randomness and deficiency. In particular,
\citet{Vidakovic1983Process} introduces a process conditional complexity and a
measure of the amount of information defined via differences of conditional
complexities. This difference-of-complexities perspective aligns naturally with
the entropy-deficiency quantities introduced later in
Sections~\ref{sec:stat_effective_uses} and \ref{sec:effective_testing}.

\section{The Adjusted Complexity $KA(x)$}
\label{sec:KA}

In this section we formalize the adjusted scale $KA(x)$ and the ratio $R(x)$, and we make precise the shell-based intuition that motivates the normalization. The key technical device is a shell counting statement (Step~3) that is converted into a Martin L\"of test in Step~4; this is the exact logical pattern we will reuse later when we replace the uncomputable $K(x)$ by computable description lengths and turn the resulting deficiency into an implementable statistic (Sections~\ref{sec:stat_effective_uses} and \ref{sec:effective_testing}). Readers primarily interested in the asymptotic theorem may skim Steps~0--3 and proceed directly to the theorem statement in Section~\ref{sec:KA_theorem}.

\subsection*{Definitions and the normalization principle}

Define the entropy-normalized Kolmogorov complexity by
\be
KA(x)=\frac{K(x)}{H(\nu_{\ell(x)})}.
\label{eq:KA_def_main}
\ee
$KA(x)$ is undefined when $H(\nu_{\ell(x)})=0$, that is, for constant words.
For nonconstant words it is finite and provides a scale-free measure of descriptive complexity.
In practice, constant words can be handled separately: they are maximally imbalanced and trivially compressible. The interesting regime is $0<p(x)<1$, where empirical entropy provides a meaningful combinatorial baseline.

Two standard baselines motivate the normalization. First, for fixed length
$\ell(x)=n$ and fixed weight $w(x)=k$, the shell
$$
S(n,k)=\{x\in\{0,1\}^n : w(x)=k\}
$$
has cardinality $\binom{n}{k}$, so the shell-typical indexing cost satisfies
$$
\log_2\binom{n}{k}
=
nH(k/n)+O(\log n),
$$
where the $O(\log n)$ term accounts for unavoidable self-delimiting overhead.
Second, for any word $x$ of length $n$, there exists a uniform coding scheme
that, given $n$, first describes the weight $w(x)$ using $O(\log n)$ bits and
then indexes $x$ within the shell $S(n,w(x))$. Consequently,
\begin{eqnarray}
K(x\mid n)
\dom
\log_2\binom{n}{w(x)}+\log n.
\label{eq:K_upper_shell}
\end{eqnarray}
Thus the quantity $H(\nu_{\ell(x)})$ provides the appropriate per-symbol
combinatorial baseline, and the adjusted complexity
$KA(x)=K(x)/H(\nu_{\ell(x)})$ rescales Kolmogorov complexity relative to the
effective size of the empirical entropy shell.

It is useful to define the dimensionless normalized ratio
\be
R(x)=\frac{K(x)}{\ell(x)\,H(\nu_{\ell(x)})},
\label{eq:R_def_main}
\ee
whenever $H(\nu_{\ell(x)})>0$. For random objects under regular measures, $R(x)$ should be close to one. Values substantially below one indicate structure beyond symbol imbalance. Later, in Sections~\ref{sec:stat_effective_uses} and \ref{sec:effective_testing}, we return to $R(x)$ and show how computable upper bounds on $K(x)$ produce effective ratios that can be calibrated into within-shell tests.

\subsection*{Exchangeable measures and the de Finetti reduction}

Let $\omega$ be Martin L\"of random under a constructive exchangeable measure $\mu$ on $\Omega$.
Exchangeability implies that $\mu$ is a mixture of i.i.d.\ Bernoulli measures \citep{deFinetti1937,DiaconisFreedman1980}, and for $\mu$--typical sequences the empirical frequency of ones converges to a limit $p\in[0,1]$.
In the constructive setting we assume (as part of the regularity hypotheses used in the main theorem) that Martin L\"of randomness with respect to $\mu$ yields an effective reduction to a Bernoulli component, namely that there exists a parameter $p$ such that $\omega$ is Martin L\"of random with respect to the Bernoulli measure $\mu_p$. This is the point at which exchangeability enters the argument, and it clarifies why the correct normalization involves the single-symbol empirical entropy $H(\nu_m(\omega))$.

\subsection*{Theorem}
\label{sec:KA_theorem}

For such a sequence $\omega$ with limiting frequency $p$,
\be
\lim_{m\to\infty}\frac{KA(\omega^{m})}{m}=1,
\qquad\text{equivalently}\qquad
\lim_{m\to\infty}\frac{K(\omega^{m})}{m}=H(p).
\label{eq:main_theorem_statement}
\ee
The proof is written to foreground a shell deficiency quantity, because the same deficiency reappears later as an effective test statistic once $K(\cdot)$ is replaced by computable coding lengths (Sections~\ref{sec:stat_effective_uses} and \ref{sec:effective_testing}).

\subsection*{Proof}
\label{sec:KA_theorem_proof}

Write $k_m=w(\omega^{m})$ and $p_m=k_m/m$. By definition, $\nu_m(\omega)=(p_m,1-p_m)$.
Since $\omega$ is Martin L\"of random under the regular exchangeable $\mu$, the empirical frequency converges,
\be
p_m\to p,
\label{eq:pm_to_p}
\ee
and therefore continuity of the binary entropy gives
\be
H(\nu_m(\omega)) \to H(p).
\label{eq:Hpm_to_Hp}
\ee
Consequently,
\be
\frac{KA(\omega^{m})}{m}
=
\frac{K(\omega^{m})}{m\,H(\nu_m(\omega))}
\label{eq:KA_over_m_ratio}
\ee
converges to $1$ if and only if
\be
\frac{K(\omega^{m})}{m}\to H(p).
\label{eq:K_over_m_target}
\ee
We prove \eqref{eq:K_over_m_target} by comparing $K(\omega^{m})$ to the shell baseline $\log_2\binom{m}{k_m}$ and showing that the per-symbol difference vanishes. The argument has two parts: a constructive upper bound on $K(\omega^{m})$ via shell coding (Step~2), and a deficiency control obtained from a counting lemma and a Martin L\"of test construction (Steps~3--4).

\medskip
\noindent
{\bf Step 0: Shells and the baseline.}
For integers $(m,k)$ define the shell (type class)
\be
S(m,k)=\{x\in\{0,1\}^{m}:w(x)=k\},
\label{eq:shell_def}
\ee
so $|S(m,k)|=\binom{m}{k}$. The quantity $\log_2\binom{m}{k}$ is the optimal indexing cost for words in $S(m,k)$, up to additive  $O(\log m)$ overhead.

\medskip
\noindent
{\bf Step 1: Stirling approximation yields the correct shell rate.}
Let $k=k_m$ and $p_m=k_m/m$. Standard Stirling bounds imply
\be
\log_2\binom{m}{k}
=
mH(p_m) + O(\log m),
\label{eq:stirling_shell}
\ee
where the remainder term is explicit and uniform across the central range of $k$.
Since $p_m\to p$, we obtain
\be
\frac{1}{m}\log_2\binom{m}{k_m} \to H(p).
\label{eq:shell_rate_to_entropy}
\ee
This is the target rate for $K(\omega^{m})/m$.

\medskip
\noindent
{\bf Step 2: An explicit upper bound on $K(\omega^{m})$ via shell coding.}
Fix $m$ and $k=k_m$. There is a computable bijection between $S(m,k)$ and $\{0,1,\dots,\binom{m}{k}-1\}$ given by lexicographic indexing. Hence there exists a prefix-free code that describes any $x\in S(m,k)$ by:
(i) a self-delimiting description of $(m,k)$ and
(ii) the index of $x$ within $S(m,k)$.
This yields a uniform bound
\be
K(x) \dom \log_2\binom{m}{k} + O(\log m),
\label{eq:K_upper_shell2}
\ee
for all words $x$ of length $m$ and weight $k$. Applying this to $x=\omega^{m}$ gives
\be
K(\omega^{m}) \dom \log_2\binom{m}{k_m} + O(\log m).
\label{eq:K_upper_on_prefix}
\ee
Dividing by $m$ and using \eqref{eq:shell_rate_to_entropy} yields
\be
\limsup_{m\to\infty}\frac{K(\omega^{m})}{m}
\le
H(p).
\label{eq:limsup_bound}
\ee

\medskip
\noindent
{\bf Step 3: Counting lemma: large deficiency inside a shell is rare.}
Define the shell deficiency of a word $x$ of length $m$ and weight $k$ by
\be
d(m,k;x)=\log_2\binom{m}{k}-K(x).
\label{eq:deficiency_def}
\ee
Fix $(m,k)$ and an integer $t\ge 1$, and consider the set
\be
A_{m,k}(t)=\{x\in S(m,k): d(m,k;x)\ge t\}.
\label{eq:Amk_def}
\ee
Then
\be
|A_{m,k}(t)| \le 2^{-t}\binom{m}{k}.
\label{eq:counting_lemma}
\ee
Indeed, the condition $d(m,k;x)\ge t$ is equivalent to $K(x)\le \log_2\binom{m}{k}-t$.
There are at most $2^{\log_2\binom{m}{k}-t}$ programs of length at most $\log_2\binom{m}{k}-t$, hence at most that many distinct outputs. Therefore at most a $2^{-t}$ fraction of the shell can have deficiency at least $t$.

\medskip
\noindent
{\bf Step 4: Turning the counting lemma into a Martin L\"of test.}
For each $m\ge 1$ and each $t\ge 1$, define the set of sequences whose prefix of length $m$ has deficiency at least $t$ by
\be
B_{m}(t)=\{\omega\in\Omega: d(m,w(\omega^{m});\omega^{m})\ge t\}.
\label{eq:Bm_def}
\ee
We now bound the probability of $B_m(t)$ under the Bernoulli measure $\mu_p$.
For any word $x$ of length $m$ and weight $k$, we have
\be
\mu_p(\Gamma_{x})=p^{k}(1-p)^{m-k}.
\label{eq:bern_cylinder_prob}
\ee
Hence, summing over $k$ and then over the atypical set $A_{m,k}(t)$ within each shell,
\begin{eqnarray}
\mu_p(B_m(t))
&=&
\sum_{k=0}^{m}\ \sum_{x\in A_{m,k}(t)} \mu_p(\Gamma_x)
\nonumber\\
&=&
\sum_{k=0}^{m} |A_{m,k}(t)|\, p^{k}(1-p)^{m-k}
\nonumber\\
&\le&
\sum_{k=0}^{m} 2^{-t}\binom{m}{k}\, p^{k}(1-p)^{m-k}
\nonumber\\
&=&
2^{-t} (p + (1-p))^m
\nonumber\\
&=&
2^{-t}.
\end{eqnarray}
Now define a global test by weighting over $m$ in a computable way, for example
\be
\mathcal{U}_t=\bigcup_{m\ge 1} B_m(t+2\log_2(m+1)).
\label{eq:Ut_def}
\ee
Using the previous bound with $t$ replaced by $t+2\log_2(m+1)$ yields
\begin{eqnarray}
\mu_p(\mathcal{U}_t)
&\le&
\sum_{m\ge 1} \mu_p\Big(B_m(t+2\log_2(m+1))\Big)
\nonumber\\
&\le&
\sum_{m\ge 1} 2^{-(t+2\log_2(m+1))}
\nonumber\\
&=&
2^{-t}\sum_{m\ge 1}(m+1)^{-2}
\;<\;
c\,2^{-t},
\label{eq:Ut_prob_bound}
\end{eqnarray}
for an absolute constant $c$. The family $\{\mathcal{U}_t\}_{t\ge 1}$ is uniformly r.e.\ (because $K(x)$ is upper semicomputable, hence the predicate $d(m,k;x)\ge r$ is effectively enumerable), and therefore constitutes a Martin L\"of test with respect to $\mu_p$.

Since $\omega$ is Martin L\"of random with respect to $\mu_p$, it avoids this test.
Hence there exists a constant $t_0$ such that for all sufficiently large $m$,
\be
d(m,k_m;\omega^{m})
<
t_0 + 2\log_2(m+1).
\label{eq:deficiency_bound}
\ee
Equivalently,
\be
\log_2\binom{m}{k_m} - K(\omega^{m}) = O(\log m).
\label{eq:log_gap_Ologm}
\ee
Dividing by $m$ gives
\be
\frac{1}{m}\Big(\log_2\binom{m}{k_m}-K(\omega^{m})\Big)\to 0.
\label{eq:gap_over_m_to_zero}
\ee

\medskip
\noindent
{\bf Step 5: conclusion.}
Combine \eqref{eq:shell_rate_to_entropy} and \eqref{eq:gap_over_m_to_zero}:
\begin{eqnarray}
\frac{K(\omega^{m})}{m}
&=&
\frac{1}{m}\log_2\binom{m}{k_m}
-
\frac{1}{m}\Big(\log_2\binom{m}{k_m}-K(\omega^{m})\Big)
\nonumber\\
&\to&
H(p)-0
=
H(p).
\end{eqnarray}
Therefore, using \eqref{eq:Hpm_to_Hp},
\be
\frac{KA(\omega^{m})}{m}
=
\frac{K(\omega^{m})}{m\,H(\nu_m(\omega))}
\to
1,
\ee
as claimed.

\medskip
\noindent
{\bf Remark.}
The core of the argument is a deficiency control: within each shell, words with a large deficiency $\log_2\binom{m}{k}-K(x)$ form an exponentially small subset, and Martin L\"of randomness rules out infinitely many large deficiencies. This difference-of-complexities logic parallels the process-based information measures developed in \citep{Vidakovic1983Process}, where information is defined via differences of conditional process complexities and is tied to randomness properties of sequences. In later sections we effectively reuse this logic by replacing $K(x)$ with computable description lengths, thereby turning the deficiency into a test statistic; see, in particular, the effective ratios in Section~\ref{sec:stat_effective_uses} and the consolidated testing rules in Section~\ref{sec:effective_testing}.

\subsection{Non-ergodic and Pathological Measures}
\label{sec:pathological}

The regularity assumptions imposed on the underlying measure $\mu$ are essential for the validity and interpretability of the limit theorem. The key point is that the baseline used in the theorem is the single-symbol empirical entropy $H(\nu_m(\omega))$, which captures marginal symbol frequencies but ignores higher-order constraints. If the generating mechanism enforces strong dependencies while preserving an apparently balanced marginal distribution, then $H(\nu_m(\omega))$ can be close to $1$ even though the true combinatorial growth rate of admissible strings is smaller. This is precisely the setting in which a higher-order normalization, based on empirical block entropies or conditional entropies, becomes necessary. This observation is echoed later when we introduce conditional empirical baselines in Section~\ref{sec:conditional_mutual}.

To illustrate the point concretely, consider a process that emits blocks $00$, $01$, and $11$ with equal probability and never produces the block $10$. More precisely, define a measure $\mu$ on $\Omega$ supported on sequences $\omega$ such that for every $k\ge 1$ the length-$2$ block $(\omega_{2k-1},\omega_{2k})$ belongs to $\{00,01,11\}$. This measure is highly constrained and is not exchangeable and not i.i.d.

For $\mu$--typical sequences, the marginal frequency of ones is still $1/2$, since each block has expected number of ones equal to $1$, so the empirical distribution satisfies
\be
\nu_m(\omega)\to(1/2,1/2),
\qquad
H(\nu_m(\omega))\to 1.
\ee
However, the number of admissible length-$m$ words is of order $3^{m/2}$ (up to a factor depending on parity), since there are $3$ choices for each disjoint pair. Thus the true combinatorial growth rate of the support is
\be
\frac{1}{m}\log_2(3^{m/2})=\frac{1}{2}\log_2 3 < 1.
\ee
For $\mu$--random sequences $\omega$ (random with respect to this constrained block measure), the Kolmogorov complexity of $\omega^{m}$ tracks the logarithm of the number of admissible prefixes, and therefore one expects
\be
\frac{K(\omega^{m})}{m}\to \frac{1}{2}\log_2 3,
\ee
while at the same time $H(\nu_m(\omega))\to 1$. Consequently,
\be
\frac{K(\omega^{m})}{m\,H(\nu_m(\omega))}
\to
\frac{(1/2)\log_2 3}{1}
=
\frac{1}{2}\log_2 3,
\ee
which is strictly less than $1$.

This example shows that the limit theorem for entropy normalization may fail when the underlying measure has dependence structure that is invisible to the single-symbol empirical distribution. It also clarifies what the regularity assumptions accomplish: in the exchangeable i.i.d.\ mixture regime, the one-bit empirical entropy is the correct baseline, whereas in constrained or non-ergodic regimes it may be necessary to replace $H(\nu_m)$ by block entropies, empirical conditional entropies, or other higher-order empirical summaries in order to obtain a stable and interpretable normalization. For this reason the conditional normalization developed later in Section~\ref{sec:conditional_mutual} is not merely a formal extension; it is a practical response to precisely this type of hidden dependence.

\section{Algorithmic Interpretation and Complexity-Theoretic Variants}
\label{sec:algo_interp}

The entropy-adjusted complexity $KA(x)$ isolates intrinsic algorithmic
structure by removing the purely combinatorial contribution arising from
symbol imbalance. For Martin L\"of random sequences generated by regular
constructive measures, the normalized ratio
\be
\frac{KA(\omega^{m})}{m}
=
\frac{K(\omega^{m})}{m\,H(\nu_m(\omega))}
\label{eq:ratio_limit_algo}
\ee
converges to one, as established in Section~\ref{sec:KA_theorem_proof}. This
alignment shows that Kolmogorov complexity achieves the optimal empirical
entropy rate dictated by the observed symbol frequencies. The present section
interprets this fact from several complementary viewpoints and clarifies how
entropy normalization fits naturally into existing algorithmic and
complexity-theoretic frameworks.

\subsection*{Universal priors and within-shell normalization}

Within the class of semi calculable measures on $\Omega$, universal
semimeasures dominate all others up to multiplicative constants and therefore
assign relatively large mass to objects of low Kolmogorov complexity
\citep{Vidakovic1988}. From this perspective, entropy normalization may be
viewed as a \emph{local} analogue of the universal prior principle, restricted
to a fixed empirical shell determined by $\nu_{\ell(x)}$.

Conditioning on the shell removes the combinatorial contribution associated
with symbol imbalance. The empirical entropy $H(\nu_{\ell(x)})$ determines the
logarithmic size of this shell, while $K(x)$ measures the descriptive cost
relative to a universal reference machine. Thus $KA(x)$ quantifies the
algorithmic deviation of $x$ from a typical shell element under the least
informative prior consistent with its empirical distribution. This viewpoint
makes explicit why deficiencies of the form
\be
\ell(x)H(\nu_{\ell(x)})-K(x)
\label{eq:deficiency_recalled_algo}
\ee
play a central role both in the proof of the main theorem
(Section~\ref{sec:KA_theorem_proof}) and in the effective testing procedures
developed later (Section~\ref{sec:effective_testing}).

\subsection*{Measure-sensitive notions of randomness}

Entropy normalization is compatible with refined notions of algorithmic
randomness that explicitly incorporate measure structure. Process-based
complexities and Schnorr-style formulations condition descriptions on
computable probability measures or generating mechanisms
\citep{Vidakovic1988}. For fixed empirical entropy, the adjusted complexity
$KA(x)$ behaves as a rescaled version of $K(x)$ and preserves all order
comparisons up to additive constants, expressed throughout this paper using
the domination relation $\preceq$ introduced in
Section~\ref{sec:preliminaries}.

At the level of infinite sequences, the asymptotic behavior of
$KA(\omega^{m})$ mirrors classical complexity-rate theorems, in which the
entropy rate of the underlying measure determines the growth rate of
algorithmic complexity. At the same time, the example in
Section~\ref{sec:pathological} demonstrates that single-symbol empirical
entropy may fail as a baseline for dependent or constrained processes. This
observation motivates the conditional normalization developed in
Section~\ref{sec:conditional_mutual}, where empirical joint structure replaces
marginal symbol frequencies as the appropriate combinatorial reference.

\subsection*{Resource-bounded variants}

From a computational standpoint, it is natural to consider resource-bounded
variants of adjusted complexity. Let $K^{t}(x)$ denote time-bounded Kolmogorov
complexity, where programs are required to halt within time $t(\ell(x))$. One
may then define
\be
KA^{t}(x)=\frac{K^{t}(x)}{H(\nu_{\ell(x)})},
\qquad
R^{t}(x)=\frac{K^{t}(x)}{\ell(x)\,H(\nu_{\ell(x)})}.
\label{eq:time_bounded_adjusted}
\ee
For fixed time bounds, these quantities are total recursive and therefore
computable. They provide a principled bridge between ideal algorithmic
descriptions and feasible coding under computational constraints. This bridge
is made explicit in Section~\ref{sec:effective_testing}, where $K(x)$ is
replaced by concrete computable surrogates $K_{\mathrm{eff}}(x)$ and the
resulting normalized ratios are used as operational test statistics.

\subsection*{Connections to MDL and universal coding}

Entropy normalization also aligns naturally with ideas from statistical model
selection and universal coding. In the minimum description length framework,
model complexity and data fit are balanced through two-part codes
\citep{Rissanen1978,BarronRissanenYu1998,BarronClarke1990}. In the present
setting, the quantity $\ell(x)H(\nu_{\ell(x)})$ represents the stochastic code
length associated with the empirical model, while $K(x)$ captures the
structural description length. Their ratio,
\be
R(x)=\frac{K(x)}{\ell(x)\,H(\nu_{\ell(x)})},
\label{eq:R_MDL}
\ee
measures algorithmic structure beyond what is implied by symbol frequencies
alone.

Universal coding schemes such as arithmetic coding achieve code lengths close
to $\ell(x)H(\nu_{\ell(x)})$ for typical sequences generated by stationary
sources. The main theorem in Section~\ref{sec:KA} shows that Kolmogorov
complexity itself matches this empirical baseline, up to $O(\log m)$ overhead
for finite prefixes, whenever the sequence is Martin L\"of random under a
regular measure. Consequently,
\be
R(\omega^{m})\to 1,
\ee
providing an algorithmic justification for entropy-normalized complexity as the
appropriate scale for assessing intrinsic randomness and structure in finite
binary data.

\section{Statistical, Applied, and Effective Uses of Adjusted Complexity}
\label{sec:stat_effective_uses}

The entropy-normalized complexity framework admits both conceptual and practical
interpretations once the abstract shell arguments of
Section~\ref{sec:KA_theorem_proof} are translated back to finite data.
A central summary statistic is the normalized randomness score
\be
R(x)=\frac{K(x)}{nH(\nu_n)}, \qquad n=\ell(x),
\label{eq:R_stat_section_clean}
\ee
which compares the descriptive cost of a word to the combinatorial baseline
implied by its empirical distribution. Values of $R(x)$ close to one indicate
that $x$ is algorithmically typical relative to its empirical shell, while
values substantially below one signal algorithmic structure beyond what symbol
frequencies alone can explain. This normalization is essential in applications
where the proportion of ones varies widely across samples, as it prevents
sparsity or imbalance from being misinterpreted as regularity.

The abstract counting lemma \eqref{eq:counting_lemma} may already be read as a
proto-statistical statement: within a fixed empirical shell, large deficiencies
are exponentially rare. The present section shows how this principle becomes an
operational tool when combined with effective approximations to Kolmogorov
complexity.

Potential applications include randomness testing, complexity-based hypothesis
testing, assessment of structure in symbolic sequences, analysis of biological
and genomic data, and evaluation of binary patterns arising from thresholding
procedures such as wavelet coefficient selection. In all these settings,
entropy normalization ensures that comparisons are made relative to the
appropriate combinatorial baseline rather than raw word length.

\subsection*{Shells and typicality revisited}

Let $x\in\{0,1\}^n$ be a finite binary word with $n=\ell(x)$ and weight $w(x)=k$.
The associated shell (type class)
\be
S(n,k)=\{x\in\{0,1\}^n:w(x)=k\}
\ee
contains exactly $\binom{n}{k}$ words, all sharing the same empirical
distribution $\nu_n=(p,1-p)$ with $p=k/n$. Consequently,
\be
\log_2|S(n,k)|=\log_2\binom{n}{k}
= nH(p)+O(\log n),
\ee
which is the finite-word analogue of the Stirling approximation
\eqref{eq:stirling_shell} used in the proof of the main theorem.

A word $x\in S(n,k)$ is called \emph{shell-typical} if its conditional Kolmogorov
complexity relative to $(n,k)$ attains the maximal order permitted by the shell,
\be
K(x\mid n,k) \asym \log_2\binom{n}{k}.
\ee
Standard counting arguments imply that shell-atypical words with substantially
smaller complexity form an exponentially small fraction of $S(n,k)$, precisely
the phenomenon quantified abstractly in \eqref{eq:counting_lemma}.

\subsection*{Why normalization matters: balanced versus imbalanced shells}

The effect of entropy normalization is illustrated by comparing shell-typical
words with different empirical distributions. Consider two words of length
$n=1000$. Let $x$ contain exactly $500$ ones arranged in a shell-typical manner,
so that $p_x=0.5$ and $H(\nu_x)=1$. Let $y$ contain exactly $100$ ones, also
arranged in a shell-typical manner, so that $p_y=0.1$ and
\be
H(\nu_y)
=
-0.1\log_2(0.1)-0.9\log_2(0.9)
\approx0.469.
\ee
Shell typicality implies
\be
K(x)\approx nH(\nu_x)=1000,
\qquad
K(y)\approx nH(\nu_y)\approx469.
\ee
Although the absolute complexities differ substantially, the adjusted scales
coincide:
\be
KA(x)\approx1000,
\qquad
KA(y)\approx1000.
\ee
The normalization removes the trivial combinatorial effect of imbalance and
isolates intrinsic descriptive complexity.

The contrast with structured words is equally instructive. Let
$z=010101\ldots01$ be the alternating word of length $n=1000$. Then
$p_z=0.5$ and $H(\nu_z)=1$, but $K(z)$ is of order $\log_2 n$ because a short
program generates the pattern. Consequently, $KA(z)$ is of order $\log_2 n$,
far below $n$. This is precisely the intended behavior: $z$ shares the same
empirical distribution as a random balanced word, yet exhibits far less
algorithmic complexity.

\subsection*{Effective surrogates and operational statistics}

Kolmogorov complexity is not computable, and therefore neither $KA(x)$ nor
$R(x)$ can be evaluated exactly. Effective surrogates are obtained by replacing
$K(x)$ with computable upper bounds. Since $K(x)$ is upper semicomputable, there
exists a computable sequence $K_t(x)$ decreasing to $K(x)$ as $t\to\infty$. For
fixed $t$, define
\be
KA_t(x)=\frac{K_t(x)}{H(\nu_n)},
\qquad
R_t(x)=\frac{K_t(x)}{nH(\nu_n)}.
\label{eq:Kt_defs_clean}
\ee

A fully effective alternative is obtained by fixing a computable lossless code
$C$ and setting $K_C(x)=L_C(x)$, the code length of $x$ under $C$. The operational
statistic
\be
R_C(x)=\frac{L_C(x)}{nH(\nu_n)}
\label{eq:RC_def_clean}
\ee
is computable and directly interpretable. The coding theorem guarantees
\be
K(x)\preceq L_C(x),
\ee
so $R_C(x)$ provides a principled upper-bound surrogate for $R(x)$.

Two explicit effective bounds are particularly instructive. The literal code
yields
\be
K_{\mathrm{len}}(x)=n,
\ee
while a combinatorial shell code that encodes $(n,w(x))$ and indexes $x$ within
$S(n,w(x))$ yields
\be
K_{\mathrm{comb}}(x)=\log_2\binom{n}{w(x)}+O(\log n).
\ee

Given any such computable upper bound $K_{\mathrm{eff}}(x)$, define the effective
normalized complexity
\be
R_{\mathrm{eff}}(x)=\frac{K_{\mathrm{eff}}(x)}{nH(\nu_n)}.
\label{eq:Reff_def_clean}
\ee

\subsection*{Deficiency-based testing}

It is often convenient to work with entropy deficiency rather than ratios.
Define
\be
d_A(x)=nH(\nu_n)-K(x),
\ee
and its effective analogue
\be
d_A^C(x)=nH(\nu_n)-L_C(x).
\ee
Fix $m\ge1$ and reject randomness whenever $d_A^C(x)\ge m$, equivalently when
\be
R_C(x)\le1-\frac{m}{nH(\nu_n)}.
\label{eq:test_threshold_clean}
\ee
This rule is the practical incarnation of the deficiency logic used in the
proof of the main theorem: compare an observed description length to the shell
baseline and flag unusually large gaps. This construction is closely related to randomness deficiency in algorithmic statistics, where typicality is quantified by the gap between a description length and an entropy-based baseline. When the baseline is taken to be the empirical entropy $nH(\nu_n)$, corresponding to the size of the empirical entropy shell, $d_A(x)$ agrees up to logarithmic terms with the Vereshchagin-Vitányi-Shen randomness deficiency framework \citep{VereshchaginVitanyi2004, VereshchaginShen2015}.

\subsection*{Illustrative finite example}

Consider the binary word
\be
x=01010001001000001010000100000100001,
\ee
of length $n=35$, containing $w(x)=9$ ones. The empirical entropy is
\be
H(\nu_n)\approx0.821,
\qquad
nH(\nu_n)\approx28.7\ \text{bits}.
\ee
The literal code yields $R_{\mathrm{len}}(x)\approx1.22$, while the
combinatorial shell code yields $R_{\mathrm{comb}}(x)\approx1.09$. Choosing
$m=5$ gives the threshold $c(5)\approx0.83$, so randomness is not rejected under
this crude effective coding scheme.

Entropy normalization ensures that this test compares words relative to the
correct combinatorial baseline and detects genuine algorithmic structure rather
than trivial symbol imbalance. The next section extends this logic by replacing
single-symbol empirical entropy with conditional and joint empirical baselines,
allowing side information and dependence structure to be incorporated in a
systematic way.

\section{Conditional and Mutual Adjusted Complexity}
\label{sec:conditional_mutual}

The entropy-normalized framework extends naturally to conditional and mutual
settings, providing a principled way to factor out empirical joint structure in
the presence of side information. As in the unconditional case, normalization
serves to separate purely combinatorial effects from genuine algorithmic
dependence. Conceptually, this section can be read as an answer to the issue raised by the pathological example in Section~\ref{sec:pathological}: if single-symbol marginals do not reflect the true constraints, one must enlarge the empirical summary used in the baseline.

Let $x,y\in\{0,1\}^n$ be finite binary words of common length $n$.
Define the empirical joint frequencies by
\be
p_{ab}=\frac{1}{n}\#\{i:x_i=a,\ y_i=b\}, \qquad a,b\in\{0,1\}.
\label{eq:joint_freqs}
\ee
Let $p_{\cdot b}=\sum_a p_{ab}$ and $p_{a\mid b}=p_{ab}/p_{\cdot b}$ whenever
$p_{\cdot b}>0$. The empirical conditional entropy is then
\be
H_{\mathrm{emp}}(X\mid Y)
=
\sum_b p_{\cdot b}
\Big(-\sum_a p_{a\mid b}\log_2 p_{a\mid b}\Big).
\label{eq:emp_cond_entropy}
\ee

We define the conditional adjusted complexity and its normalized ratio by
\be
KA(x\mid y)=\frac{K(x\mid y)}{H_{\mathrm{emp}}(X\mid Y)},
\qquad
R(x\mid y)=\frac{K(x\mid y)}{nH_{\mathrm{emp}}(X\mid Y)},
\label{eq:conditional_defs}
\ee
whenever $H_{\mathrm{emp}}(X\mid Y)>0$. The associated conditional entropy
deficiency
\be
d_A(x\mid y)=nH_{\mathrm{emp}}(X\mid Y)-K(x\mid y)
\label{eq:conditional_deficiency}
\ee
is a special case of Levin’s randomness deficiency \citep{Levin1974}.

Here and throughout, the symbols $X$ and $Y$ do not denote abstract random
variables generated by an underlying stochastic process. Instead, they refer to
the empirical random variables induced by the fixed finite words
$x=(x_1,\dots,x_n)$ and $y=(y_1,\dots,y_n)$: an index $i$ is drawn uniformly from
$\{1,\dots,n\}$ and $X(i)=x_i$, $Y(i)=y_i$. All entropies are therefore computed
with respect to the empirical joint distribution determined by $(x,y)$. This
convention is standard in the theory of types and empirical distributions
\citep{CoverThomas2006} and is used explicitly when relating complexity to
entropy rates \citep{LiVitanyi2008}.

Dependence between finite objects can also be studied through algorithmic mutual
information. The algorithmic mutual information between two finite words $x$ and
$y$ is defined, up to an additive constant, by
\be
I(x:y) \asymp K(x)+K(y)-K(x,y),
\label{eq:alg_mutual_info}
\ee
and measures the amount of algorithmic information shared by $x$ and $y$
\citep{Kolmogorov1965,Levin1974,GacsTrompVitanyi2001}. From an empirical
perspective, dependence is reflected in joint symbol frequencies. Letting $X$
and $Y$ denote the empirical random variables induced by $(x_i,y_i)$, the
empirical mutual entropy is
\be
I_{\mathrm{emp}}(X;Y)
=
H_{\mathrm{emp}}(X)+H_{\mathrm{emp}}(Y)-H_{\mathrm{emp}}(X,Y).
\label{eq:emp_mutual_entropy}
\ee

We define the mutual adjusted complexity by
\be
KA(x:y)=\frac{I(x:y)}{I_{\mathrm{emp}}(X;Y)},
\label{eq:mutual_adjusted}
\ee
whenever $I_{\mathrm{emp}}(X;Y)>0$. This normalization factors out the purely
statistical component of dependence and isolates algorithmic dependence beyond
empirical correlation. When $x$ and $y$ are typical samples from a joint i.i.d.\
source, $I(x:y)$ concentrates around $nI_{\mathrm{emp}}(X;Y)$ and the adjusted
mutual complexity remains close to one. In contrast, deterministic or structural
relationships may yield large $I(x:y)$ even when empirical mutual entropy is
small, resulting in large values of $KA(x:y)$. Thus $KA(x:y)$ provides a
scale-free measure of algorithmic dependence, paralleling the role of adjusted
complexity in the unconditional and conditional settings.

We now illustrate the conditional normalization on a concrete finite example,
parallel to the unconditional case considered earlier in Section~\ref{sec:stat_effective_uses}. Let
\be
x=01010001001000001010000100000100001
\label{eq:x_example}
\ee
be the same binary word of length $n=35$. As before, $w(x)=9$ and
\be
H(\nu_n)\approx0.82,
\qquad
nH(\nu_n)\approx28.8\ \text{bits},
\ee
so that relative to its unconditional shell, $x$ does not exhibit substantial
regularity beyond symbol imbalance.

Introduce side information by letting
\be
y_0=00100011
\label{eq:y0_example}
\ee
and extending it periodically, truncating to length $n=35$ to obtain a word
$y$. Forming the coordinatewise pairs $(x_i,y_i)$ yields the empirical joint
counts shown in Table~\ref{tab:jointxy}.

\begin{table}[h]
\centering
\begin{tabular}{c|cc|c}
$x\backslash y$ & 0 & 1 & total \\
\hline
0 & 19 & 7 & 26 \\
1 & 6  & 3 & 9 \\
\hline
total & 25 & 10 & 35
\end{tabular}
\caption{Empirical joint counts for $(x,y)$.}
\label{tab:jointxy}
\end{table}

From Table~\ref{tab:jointxy}, the empirical conditional probabilities are
\be
P(X=1\mid Y=0)=\frac{6}{25},
\qquad
P(X=1\mid Y=1)=\frac{3}{10},
\ee
with corresponding conditional entropies
\be
H(X\mid Y=0)\approx0.79,
\qquad
H(X\mid Y=1)\approx0.88.
\ee
Weighting by $P(Y=0)=25/35$ and $P(Y=1)=10/35$ gives
\be
H_{\mathrm{emp}}(X\mid Y)
=
\frac{25}{35}H(X\mid Y=0)+\frac{10}{35}H(X\mid Y=1)
\approx0.47.
\ee
Thus the conditional entropy baseline satisfies
\be
nH_{\mathrm{emp}}(X\mid Y)\approx16.5\ \text{bits},
\qquad
nH(\nu_n)\approx28.8\ \text{bits}.
\ee

This reduction reflects the fact that conditioning on $y$ restricts the
combinatorial shell. The coordinates are partitioned into those with $Y=0$ and
$Y=1$, and within each subset the admissible patterns for $x$ are governed by
the corresponding conditional types. The quantity $nH_{\mathrm{emp}}(X\mid Y)$
is therefore the logarithmic size of the conditional shell of words compatible
with the observed joint structure.

A natural effective conditional description length is obtained by coding $x$
within this conditional shell, for example by separately indexing the locations
of ones among the $Y=0$ and $Y=1$ coordinates. This yields an effective upper
bound $K_{\mathrm{eff}}(x\mid y)$ and the corresponding ratio
\be
R_{\mathrm{eff}}(x\mid y)
=
\frac{K_{\mathrm{eff}}(x\mid y)}{nH_{\mathrm{emp}}(X\mid Y)}
\approx1.09.
\ee

The qualitative conclusion mirrors the unconditional case. Relative to the
conditional combinatorial baseline induced by the side information $y$, the
word $x$ does not exhibit substantial additional regularity. If $x$ were nearly
determined by $y$, then $H_{\mathrm{emp}}(X\mid Y)$ would be close to zero and a
conditional shell code would achieve a much shorter description length, driving
$R_{\mathrm{eff}}(x\mid y)$ far below one. Conditional entropy normalization
thus provides a refined within-shell comparison that separates genuine
algorithmic structure from statistical constraints imposed by side information. This conditional baseline is exactly what is used in the conditional deficiency in the following section.

\section{Effective Versions and Testing}
\label{sec:effective_testing}

Kolmogorov complexity is not computable, and therefore neither the adjusted
complexity $KA(x)$ nor its conditional and mutual variants can be evaluated
exactly. Nevertheless, the entropy-normalized framework developed in the
preceding sections admits natural and practically meaningful \emph{effective}
counterparts obtained by replacing $K(\cdot)$ with computable upper bounds.
These effective versions preserve the role of the empirical entropy baseline
and lead directly to implementable procedures for testing randomness and
detecting structure in finite data.

From a logical standpoint, this section is the operational analogue of the proof
strategy used in Section~\ref{sec:KA_theorem_proof}. There, a deficiency inside
each combinatorial shell was used to construct a Martin L\"of test and to prove
the asymptotic limit theorem. Here, the same deficiency idea is reused with
explicit, computable description lengths, turning the abstract argument into a
concrete statistical device.

Let $K_{\mathrm{eff}}(x)$ be any total recursive function that dominates
Kolmogorov complexity up to an additive constant, in the sense that
\begin{eqnarray}
K(x) \preceq K_{\mathrm{eff}}(x).
\label{eq:Keff_dominates_K_clean}
\end{eqnarray}
The simplest example is the literal code,
\begin{eqnarray}
K_{\mathrm{len}}(x)=\ell(x)=n,
\end{eqnarray}
which outputs the bits of $x$ verbatim and requires no modeling assumptions.
Although crude, it provides a universal baseline.

A more informative effective bound exploits the shell structure introduced
earlier. One first encodes $n=\ell(x)$ and the weight $w(x)$, and then indexes
$x$ among the $\binom{n}{w(x)}$ words in the corresponding shell. This yields
\begin{eqnarray}
K_{\mathrm{comb}}(x)
=
\log_2\binom{n}{w(x)}+O(\log n),
\end{eqnarray}
a form studied explicitly in the context of effective combinational complexity
by \citet{StojanovicVidakovic1987}. More generally, any fixed lossless
compression scheme $C$ induces a computable description length $L_C(x)$
satisfying
\begin{eqnarray}
K(x) \preceq L_C(x),
\end{eqnarray}
where the domination constant depends only on the chosen code. Such coding
lengths play a central role in minimum description length theory and universal
coding \citep{Grunwald2007}.

Given a chosen effective bound $K_{\mathrm{eff}}(x)$, we define the effective
normalized adjusted complexity by
\begin{eqnarray}
R_{\mathrm{eff}}(x)
=
\frac{K_{\mathrm{eff}}(x)}{nH(\nu_n)},
\qquad n=\ell(x),
\label{eq:Reff_testing_clean}
\end{eqnarray}
whenever $H(\nu_n)>0$. In the conditional and mutual settings, the same
construction applies with the empirical baselines
$nH_{\mathrm{emp}}(X\mid Y)$ and $nI_{\mathrm{emp}}(X;Y)$ introduced in
Section~\ref{sec:conditional_mutual}. In all cases, the denominator represents
the logarithmic size of the relevant empirical shell, while the numerator is a
computable description length. Thus $R_{\mathrm{eff}}(x)$ compares the observed
descriptive cost of $x$ to the cost predicted by its empirical entropy. For
shell-typical words and reasonably efficient coding bounds,
$K_{\mathrm{eff}}(x)$ is close to $nH(\nu_n)$ and $R_{\mathrm{eff}}(x)$ is near
one, whereas values substantially below one indicate algorithmic structure
beyond symbol imbalance.

It is often convenient to work with the associated entropy deficiency,
\begin{eqnarray}
d_A^{\mathrm{eff}}(x)
=
nH(\nu_n)-K_{\mathrm{eff}}(x),
\label{eq:eff_deficiency_clean}
\end{eqnarray}
which is an effective analogue of Levin-type randomness deficiency specialized
to empirical Bernoulli shells. Fix an integer $m\ge1$ and consider the rejection
region
\begin{eqnarray}
\mathcal{R}_m
=
\{x:d_A^{\mathrm{eff}}(x)\ge m\}.
\label{eq:reject_region_clean}
\end{eqnarray}
By the same counting argument used in the proof of the main theorem, among all
words of length $n$ with a fixed empirical distribution, at most
$2^{nH(\nu_n)-m}$ belong to $\mathcal{R}_m$. Consequently, under i.i.d.\
Bernoulli measures consistent with the empirical distribution, the probability
of rejection is bounded above (up to constants depending on the chosen coding
bound) by a multiple of $2^{-m}$. The parameter $m$ therefore plays the role of a
significance level: rejecting when $d_A^{\mathrm{eff}}(x)\ge m$ controls false
positives independently of symbol imbalance. This mirrors, at the effective
level, the probability bound \eqref{eq:Ut_prob_bound} obtained in the
Martin--L\"of test construction of Section~\ref{sec:KA_theorem_proof}.

The same mechanism extends directly to conditional and mutual adjusted
complexities. A computable bound $K_{\mathrm{eff}}(x\mid y)$ yields the
conditional deficiency
\begin{eqnarray}
d_A^{\mathrm{eff}}(x\mid y)
=
nH_{\mathrm{emp}}(X\mid Y)-K_{\mathrm{eff}}(x\mid y),
\end{eqnarray}
leading to tests for algorithmic structure in $x$ beyond what is explained by
the side information $y$. Likewise, replacing algorithmic mutual information
$I(x:y)$ by an effective upper bound produces practical criteria for detecting
algorithmic dependence beyond empirical correlation.

From the MDL perspective, the quantity $nH(\nu_n)$ represents the stochastic
code length associated with the empirical model, while $K_{\mathrm{eff}}(x)$
represents an explicit structural description length. Entropy normalization
therefore enforces a clean separation between empirical fit and descriptive
structure, in the spirit of two-part MDL codes
\citep{BarronRissanenYu1998,Grunwald2007}. Relative to earlier effective
combinational measures \citep{StojanovicVidakovic1987}, the present formulation
places empirical entropy at the center of the normalization and integrates
unconditional, conditional, and mutual notions into a single coherent and
operational framework.

\section{Discussion and Conclusions}
\label{sec:discussion}

The entropy-normalized complexity framework developed in this paper is guided by
a simple organizing principle: descriptive structure should be assessed relative
to the appropriate empirical combinatorial baseline. Kolmogorov complexity
measures intrinsic algorithmic description length, while empirical entropy
quantifies the size of the combinatorial class compatible with observed symbol
frequencies. Normalizing complexity by empirical entropy separates these two
effects and clarifies how much structure remains once trivial imbalance has been
factored out.

This viewpoint provides a finite, local counterpart to the classical links
between Shannon entropy, universal priors, and algorithmic complexity
\citep{Solomonoff1964,Solomonoff1964b,Chaitin1975ProgramSize}. In particular, it
makes explicit that entropy alone cannot capture algorithmic structure
\citep{Kramosil1982}: two words with identical empirical entropy may differ
radically in descriptive complexity, and conversely, raw complexity values are
not directly comparable across different empirical distributions. The adjusted
quantities $KA(x)$ and $R(x)$ resolve this mismatch by placing description length
on the correct empirical scale.

Several extensions and refinements suggest themselves naturally. The present
development focuses on binary words and single-symbol empirical entropy, but the
normalization principle extends immediately to larger alphabets. More
substantively, one may replace $H(\nu_n)$ by block entropies, empirical
conditional entropies, or other summaries that capture higher-order dependence.
Such refinements are essential when symbol-level marginals do not reflect the
true combinatorial constraints of the generating mechanism. The pathological
example in Section~\ref{sec:pathological} illustrates this vividly: when strong
dependencies are invisible at the marginal level, a higher-order empirical
baseline is required. The conditional normalization developed in
Section~\ref{sec:conditional_mutual} can be viewed as a first step in this
direction.

The framework also aligns naturally with ideas from algorithmic statistics.
Here, the empirical distribution plays the role of a coarse model, and the
associated deficiency measures residual algorithmic irregularity relative to
that model. A refined treatment could place empirical, block, or conditional
models into an explicit model class and interpret entropy-normalized
deficiencies as stochasticity indices, thereby connecting more directly to
minimal sufficient statistics and classical notions of randomness deficiency.
The conditional and mutual constructions already point toward such a synthesis.

Computational considerations raise further questions. Resource-bounded variants
of adjusted complexity, introduced in Section~\ref{sec:algo_interp}, are
especially relevant for effective testing and for understanding how feasible
coding constraints affect structure detection. While Section~\ref{sec:effective_testing}
shows how computable description lengths lead to implementable tests with
controlled false-positive rates, systematic finite-sample calibration and power
analysis under simple null models remain open problems. Establishing such
calibration would strengthen the bridge between algorithmic randomness theory
and practical statistical methodology.

From an applied perspective, entropy normalization is particularly natural in
settings where imbalance or sparsity is intrinsic rather than exceptional. In
multiresolution representations, such as wavelet skeletons, thresholded
coefficient patterns, or sparse symbolic encodings, imbalance is the default.
Raw complexity measures tend to conflate sparsity with structure, whereas
entropy-normalized complexity provides a principled way to compare descriptive
content across scales. Extending the present framework to multiscale or
hierarchical settings may therefore yield useful tools for analyzing structured
signals and symbolic data.

Finally, a quantum analogue suggests itself. In Schumacher’s theory of quantum
data compression, von Neumann entropy governs the compressibility of quantum
ensembles \citep{Schumacher1995QuantumCoding}. A quantum-adjusted complexity
would require replacing classical empirical entropy by state entropy and
measuring algorithmic structure relative to density operators rather than
symbol frequencies. Such a notion could separate structural quantum correlations
from stochastic noise in mixed or entangled states, potentially strengthening
conceptual links between algorithmic information theory, quantum coding, and
quantum statistical inference.

In summary, entropy-normalized Kolmogorov complexity provides a flexible and
unifying framework that links algorithmic information, empirical entropy, and
statistical modeling. By explicitly factoring out the combinatorial contribution
of empirical distributions, it yields quantities that are both theoretically
interpretable and operationally meaningful. The resulting perspective clarifies
the role of complexity in randomness, dependence, and structure detection, and
opens several avenues for further theoretical development and applied use.

\medskip

\noindent
{\bf Acknowledgment.~} The author acknowledges the partial support of the H.\ O.\ Hartley Chair Foundation and NSF Award 2515246 at Texas A\&M University.

\bibliography{references}

\end{document}